\documentclass[twocolumn,showpacs,preprintnumbers,amsmath,amssymb]{revtex4}
\usepackage{amssymb}     
\usepackage{graphicx}    
\usepackage{dcolumn}
\usepackage{bm}
\usepackage{color}
\input{colordvi.tex}

\begin{document}

\title{Systematic Study of Gravitational Waves from Galaxy Merger}

\author{Takahiro Inagaki$^1$, Keitaro Takahashi$^1$, Shogo Masaki$^1$, and
Naoshi Sugiyama$^{1,2,3}$}

\affiliation{
$^1$Department of Physics and Astrophysics, Nagoya University,
Nagoya 464-8602, Japan; inagaki@a.phys.nagoya-u.ac.jp\\
$^2$Institute for the Physics and Mathematics of the Universe (IPMU),
The University of Tokyo, Kashiwa, Chiba, 277-8568, Japan\\
$^3$Kobayashi-Maskawa Institute for the Origin of Particles and the Universe,
Nagoya University, Nagoya 464-8602, Japan}

\date{\today}

\begin{abstract}
A systematic study of gravitational waves from galaxy mergers, 
through $N$-body simulations, was performed. In particular,
we investigated the relative importance of galaxy components
(disk, bulge and halo) and effects of initial relative velocity,
relative angular momentum and mass ratio of the galaxies.
We found that the features of light curve of gravitational
waves, such as peak width and luminosity, are reliably simulated
with particle numbers larger than $\sim 10^4$. Dominant
contribution to gravitational wave emission came from the halo
component, while peak luminosity amounted to $10^{31}~{\rm erg/sec}$
for the collision of two halos with mass
$3.8 \times 10^{12} h^{-1}M_{\odot}$.  We also found that
the initial relative velocity in the direction of the initial
separation did not significantly affect gravitational wave emission,
while the initial relative angular momentum broadened the peak width
and suppressed the luminosity. Mass dependence of the peak
luminosity was also investigated, and we obtained evidence that
the luminosity is
proportional to the cubic mass when the scaling relation is satisfied. 
This behavior was considered by a simple analysis.
\end{abstract}


\maketitle

\section{Introduction}
\label{introduction}

Structure formation of the universe proceeds through gravitational
interaction from tiny density fluctuations. In this process,
gravitational waves with cosmological scales are expected to
be produced. In \cite{2004PhRvD..69f3002M,2007PhRvD..75l3518A,
2007PhRvD..76h4019B}, generation of gravitational waves
was studied as a second-order effect of cosmological perturbations.
It was shown that the density of gravitational wave background,
$\rho_{\rm GW}$, amounts to
$\rho_{\rm GW}/\rho_{\rm c} \sim 10^{-20} - 10^{-15}$
for a wide frequency range, where $\rho_{\rm c}$ is
the critical density. At smaller scales, gravitational waves,
radiated through the formation of dark matter halo, were studied
in \cite{2006PhRvD..73f3503C} and the density was estimated
as $\rho_{\rm GW}/\rho_{\rm c} \sim 10^{-20}$ at frequencies
$10^{-18} - 10^{-17}~{\rm Hz}$. Thus, these processes of structure
formation are imprinted on the gravitational wave background.

Although such gravitational wave background of cosmological
scales is inaccessible to direct detection, it could be
probed by measuring the B-mode polarization of the cosmic
microwave background. In actuality, several missions suited
for this aim have been planned; for example,
LiteBIRD, QUIET, POLARBeaR (see, http://cmbpol.kek.jp/index-e.html),
and Cosmic Inflation Probe (see, http://www.cfa.harvard.edu/cip/).
Their primary purpose is to detect primordial gravitational waves
generated during inflation. However, it was reported
in \cite{2006PhRvD..73f3503C} that gravitational waves from
halo formation could be dominant if the energy scale of inflation
is below $\sim 10^{15}~{\rm GeV}$.

In this paper, we investigate galaxy merger as another process which
produces gravitational waves of cosmological scales.
In the hierarchical model of structure formation,
it is argued that low-mass dark matter halos repeatedly merge
with each other to form more massive halos 
(see, e.g., \cite{1993MNRAS.262..627L,1994MNRAS.271..676L}).
Because halo merger is a highly nonlinear process which involves
huge masses, it is expected that substantial amounts of
gravitational waves are emitted. In \cite{2007PhRvD..75j4008Q},
gravitational waves from galaxy merger are calculated with $N$-body
simulations. These researchers 
studied several typical configurations of galaxy
merger and estimated the luminosity to be of order
$10^{33}~{\rm erg~sec^{-1}}$.

By extending this previous study
\cite{2007PhRvD..75j4008Q}, we perform a systematic study of
gravitational waves from galaxy mergers through $N$-body simulations
with Gadget2 \cite{2001NewA....6...79S,2005MNRAS.364.1105S}. 
While in \cite{2007PhRvD..75j4008Q} initial
condition was generated by an open code ``GalactICS''
\cite{1995MNRAS.277.1341K} and Jaffe model
\cite{1983MNRAS.202..995J}, here we use a halo adopted
from a cosmological simulation as well as ones generated by
``GalactICS''. The former has a realistic density structure
which is well approximated by the Navarro-Frenk-White (NFW)
profile \cite{1996ApJ...462..563N}. On the other hand,
the latter describe disk galaxies with three components,
disk, bulge and halo. With these initial conditions,
we investigate the relative importance of galaxy components
(disk, bulge and halo) and the effects of initial velocity,
relative angular momentum and mass ratio of the halos.
Also we check the dependence of the results on simulation
resolution by varying the particle number. This study can be
regarded as the first step toward evaluating the gravitational
wave background from galaxy merger.

The paper is organized as follows. In section \ref{method},
we explain our method of calculation, including the preparation
of the initial conditions. In section \ref{results}, the results
are presented. Finally section \ref{summary} is devoted to
the summary of our study and discussion including our future plans.

\section{Method}
\label{method}

We follow the merging process of two galaxies using
the parallel $N$-body solver Gadget2
\cite{2001NewA....6...79S,2005MNRAS.364.1105S}
in its Tree-PM mode. The gravitational softening parameter
is set to around $1/60$ of the tidal radius of galaxies.
We use two types of initial conditions (type A and B).
Type A is prepared by generating galaxies with ``GalactICS''
while a realistic halo is adopted from a cosmological simulation
for type B. Detailed description of the initial condition
will be given in the next section.

Based on the $N$-body simulation, we compute the amplitude
and luminosity of the gravitational waves.
Here, we neglect the effect of cosmological expansion
because the initial separation of galaxies is taken to be
$\sim 1~h^{-1}{\rm Mpc}$ and the ratio of acceleration
due to cosmic expansion and gravitational force between
galaxies is of order $0.1$. Further, the velocity dispersion of
stars in each galaxy and the relative velocity between two
galaxies are nonrelativistic ($v/c \leq 10^{-3}$) so that
the slow-motion approximation is valid. By denoting the deviation
of the metric from Minkowski spacetime as $h_{\mu\nu}$,
and taking the transverse-traceless (TT) gauge, the spatial
components of $h_{\mu\nu}^{TT}$ are given by the quadrupole
formula as
\begin{equation}
h_{ij}^{TT} =
\frac{2G}{c^4 r} \ddot{I}_{ij}(t-r/c),
\label{eq:hij}
\end{equation}
where $G$ is the gravitational constant and $r$ is the distance
from the observer to the merging galaxies. Here $I_{ij}$ is
the components of the traceless internal tensor given by
\begin{align}
I_{ij} =
\sum _k m^{(k)}
\left( x_i^{(k)} x_j^{(k)} - \frac{1}{3} \delta_{ij}{x^{(k)}}^2
\right)
\label{Iij}
\end{align}
where $x_i^{(k)}$ represents the position of $k$-th particle.
Later we will set the initial separation of two galaxies
in the direction of $x$.


The luminosity of gravitational waves is given by
\begin{equation}
L_{\rm GW}
= \left( \frac{dE}{dt} \right) _{\rm GW}
= \frac{G}{5c^5} \sum _{ij} \dddot{I}_{ij} \dddot{I}_{ij}
~~~ {\rm [erg~s^{-1}]}.
\label{LGW}
\end{equation}
It should be noted here that 
this formula needs third-order time derivatives
which may cause a substantial numerical error. 
We will estimate the numerical error
by three different calculational procedures
in section \ref{basic}.

\section{Results}
\label{results}

In this section, we present the results of our numerical
simulations. In subsection \ref{basic} we generate initial
conditions using ``GalactICS'' and investigate
the relative importance of the galaxy components
(halo, disk and bulge). Also we check the convergency of
the results by varying the particle number.
In subsection \ref{NFW}, we use initial conditions
adopting a dark halo from a cosmological simulation.
Finally we study the effects of the initial relative velocity,
relative angular momentum and mass (ratio).

\subsection{Simulations with Type A Initial Conditions}
\label{basic}

\begin{table}[t]
\caption{Models used in this subsection.
Column(1):model name.
Column(2):total mass.
Column(3):mass of halo.
Column(4):mass of disk.
Column(5):mass of bulge.
Column(6):disk scale radius.
Column(7):tidal radius.}
\label{tb:dbh}
\begin{tabular}{ccccccc}
\hline\hline
model 
& $M_{\rm tot}~(10^{10}~M_\odot)$
& $M_{\rm h}$
& $M_{\rm d}$
& $M_{\rm b}$
& $r_{\rm d}~({\rm kpc})$
& $r_{\rm t}~({\rm kpc})$ \\
\hline
model-0 & 31.1 & 26.5 & 3.9 & 0.7  & 4.5 & 56.5\\
model-A & 31.9 & 25.0 & 4.4 & 2.2  & 4.5 & 98.9 \\
model-D & 195.6 & 188.9 & 4.5 & 2.2  & 4.5 & 330.7 \\
\hline
\end{tabular}
\end{table}

First, let us describe the initial conditions.
In this subsection \ref{basic} we use initial conditions
generated by ``GalactICS'' \cite{1995MNRAS.277.1341K}.
This is based on a semi-analytic model for the phase-space
distribution function of an axisymmetric disk galaxy,
and outputs the distribution functions of a disc, bulge and
halo from input parameters such as total mass and angular
momentum. One of the advantages of using ``GalactICS'' is
that it can create a disk galaxy with halo, disk and bulge
of arbitrary mass self-consistently.
We will see that the dominant contribution to gravitational
wave emission comes from the halo component. Another advantage
is that we can choose the number of particles arbitrarily
so that we can study the effect of resolution on gravitational
wave emission.

\begin{figure}[t]
\includegraphics[width=70mm]{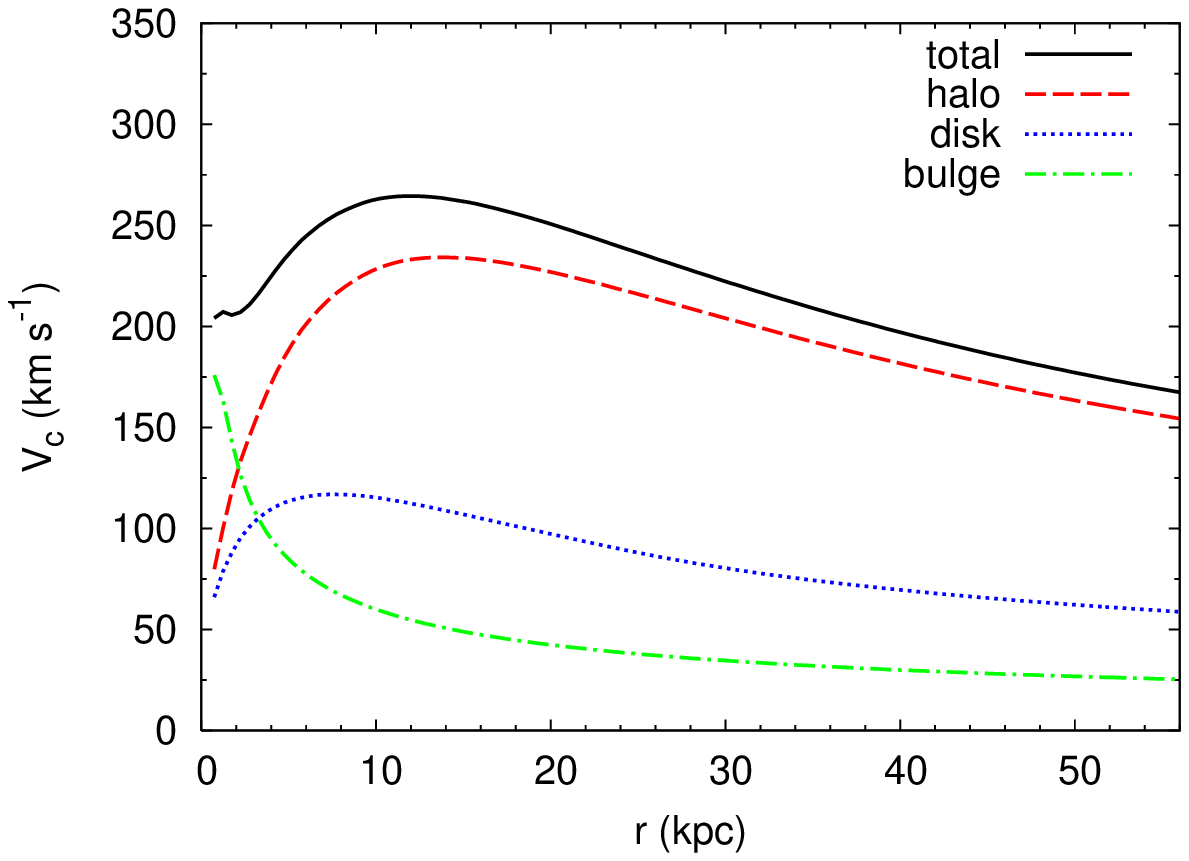}\\
\includegraphics[width=70mm]{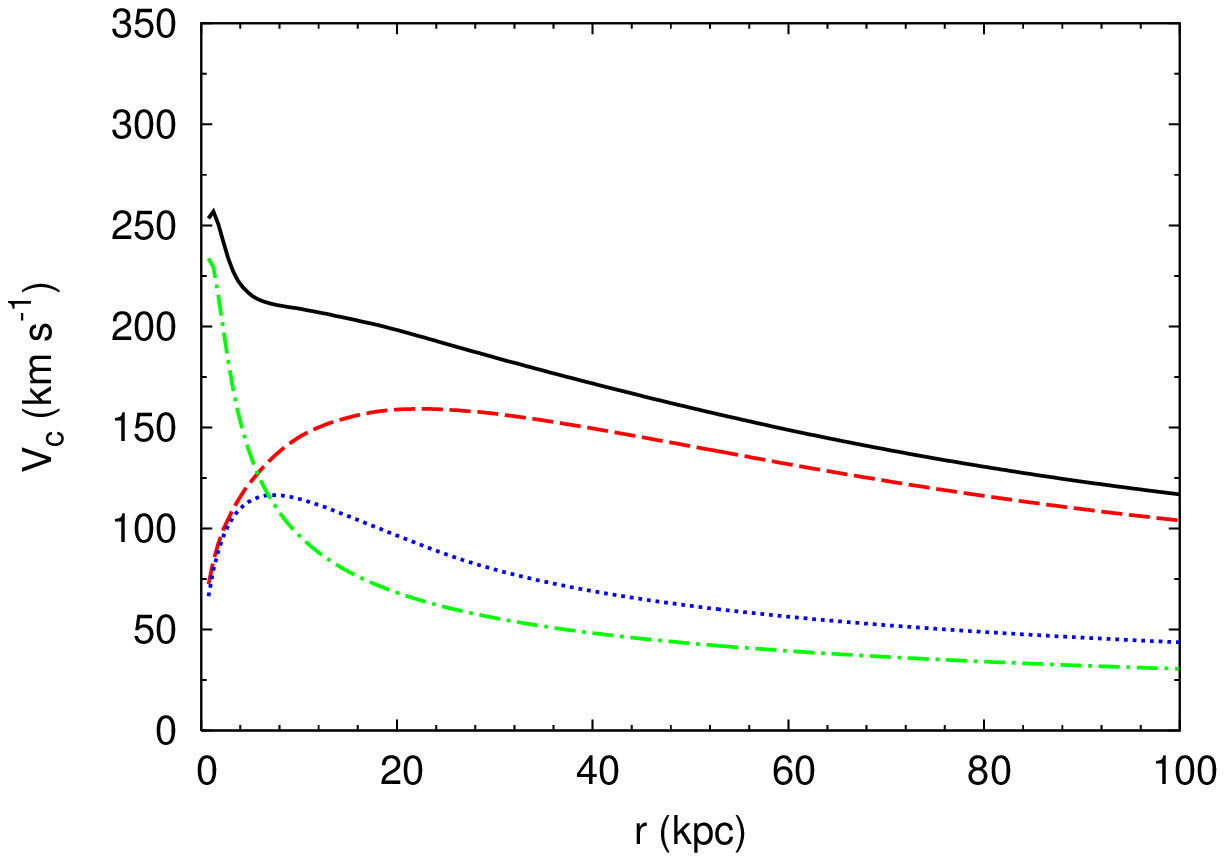}\\
\includegraphics[width=70mm]{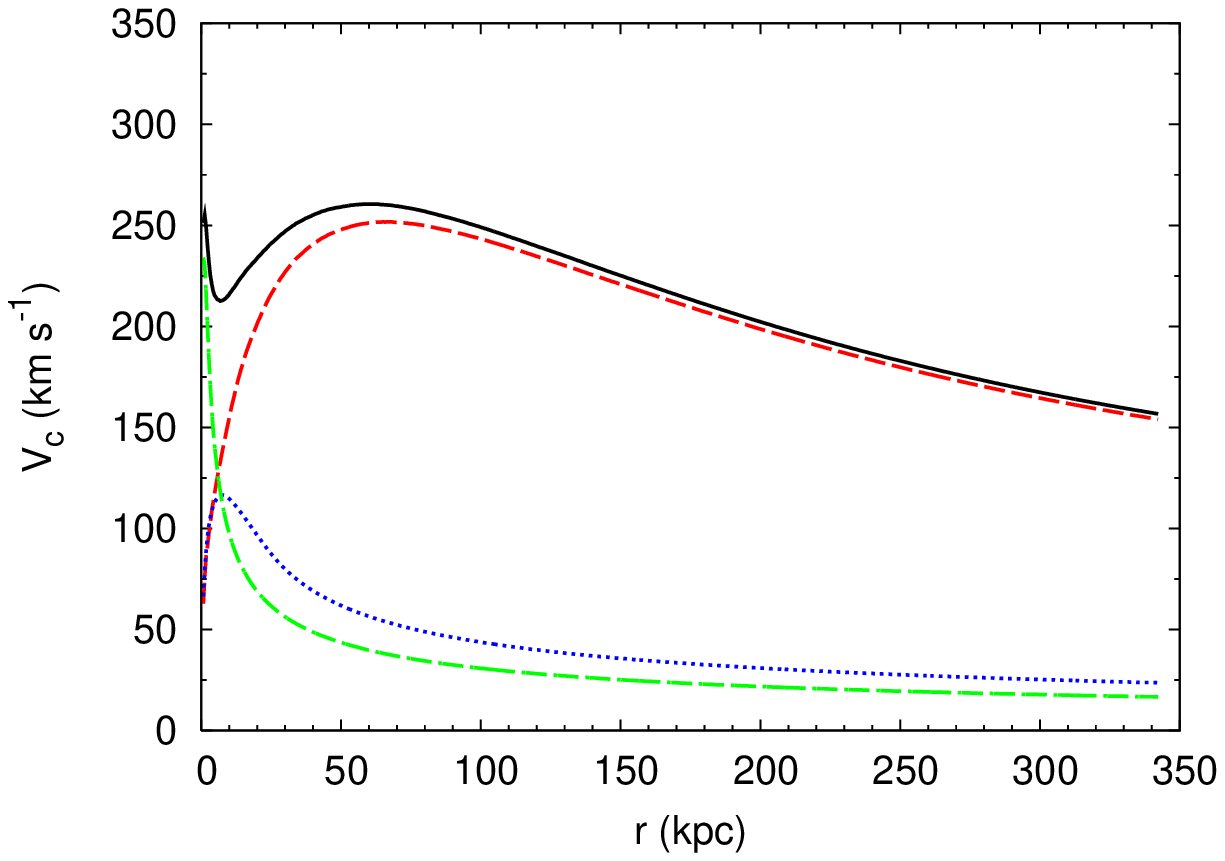}
\caption{
Rotation curve ,  $V_c(r)=\sqrt{GM(r)/r}$, 
for halo+disk+bulge (solid), halo (dashed), disk (dotted), 
and bulge (dash-dotted) for a range from galactic center to $r_{\rm t}$.
Top: model-0. Middle: model-A. Bottom: model-D.
}
\label{fig:density-0}
\end{figure}

We prepared three initial conditions whose parameters are shown
in TABLE \ref{tb:dbh}. Here $r_{\rm d}$ and $r_{\rm t}$ are
the scale radius of the disk and the tidal radius, which is
the outermost radius of the halo \cite{1995MNRAS.277.1341K},
respectively. The model-0 is similar to the disk galaxy adopted
in \cite{2007PhRvD..75j4008Q}. As we can see in Fig.
\ref{fig:density-0}, the rotation curve is decreasing outward
within $30~{\rm kpc}$, which means that the halo component
is relatively concentrated. In this sense, this model is not so
realistic. To study the effect of concentration on gravitational-wave
emission, we consider model-A, which has almost the same mass
as model-0 but a flatter rotation curve (Fig. \ref{fig:density-0}).
Further, model-D is considered to show the effect of mass and
halo extension. Model-D has a similar mass and structure as our galaxy
and the central density profile is similar to model-A.
It should be noted that model-A and model-D are
identical to model A and model D in \cite{1995MNRAS.277.1341K}.
In this subsection, we will mainly use model-0 and, at the end,
make a comparison between model-0 and model-A.

The particle number per galaxy is set to $5.4 \times 10^5$
for all models in this section. Then we simulate the head-on merger
of two identical galaxies with initial relative velocity
in the $x$-direction which corresponds to the kinetic energy
equal to potential energy.

\begin{figure}[t]
\includegraphics[width=77mm]{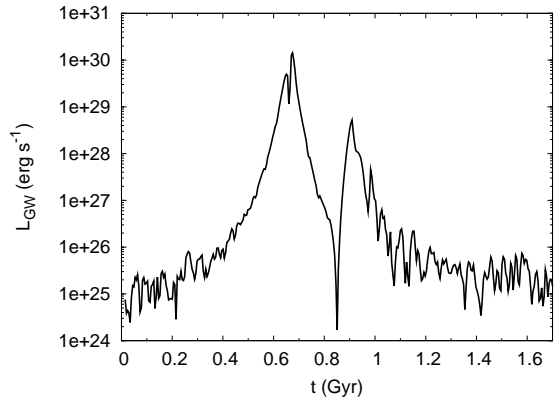}
\caption{Time evolution of luminosity of gravitational waves for model-0.}
\label{fig:hdb0}
\end{figure}

Fig. \ref{fig:hdb0} shows the time evolution of the luminosity
of gravitational wave for model-0. We can see two major
peaks which correspond to the collisions of two galaxies.
These gradually relax into a single galaxy and gravitational-wave
emission becomes much less effective after $t = 1.2~{\rm Gyr}$.
The peak width, $\sim 0.1~{\rm Gyr}$, reflects the dynamical
timescale of the galaxies. The peak luminosity reaches
$1.4 \times 10^{30}~{\rm erg/sec}$ and the total emitted energy is
about $\sim 10^{46}~{\rm erg}$. As we will see below, the steady
emission after $t = 1.2~{\rm Gyr}$ is substantially affected by
the simulation resolution and would be considered as noise
within the simulation. Therefore, there is a possibility that
more peaks do exist below this noise level. Nevertheless,
it is robust that the luminosities of the third and later peaks
are less than $\sim 10^{26}~{\rm erg/sec}$ and make negligible
contributions to the total luminosity.

\begin{figure}[t]
\includegraphics[width=77mm]{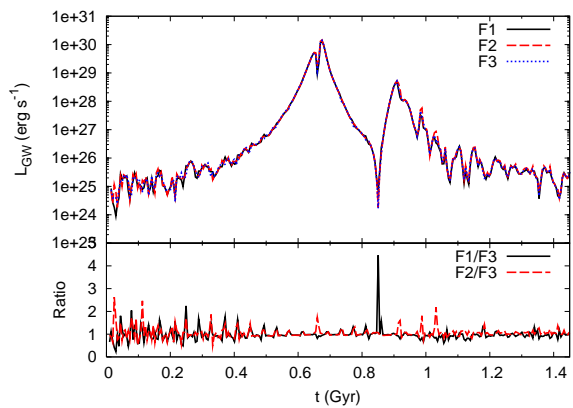}
\caption{
Top: comparison of gravitational wave luminosities calculated
with three different approaches for model-0. See text for detail.
Bottom: ratios of luminosity, F1/F3 and F2/F3.}
\label{fig:hdb3}
\end{figure}

As we saw in Eq. (\ref{LGW}), we need
the third-order derivatives
of $I_{ij}$ to calculate gravitational wave luminosity.
However, if we calculate these by differentiating $I_{ij}$ three times
numerically, it may cause a substantial numerical error.
Because Gadget2 outputs $\ddot{x}_i$ as well as $x_i$ and
$\dot{x}_i$, it is possible to obtain $\dddot{I}_{ij}$ with just
one numerical differentiation, allowing for the reduction of 
numerical error.
To aide in estimating the error concerning the numerical differentiation,
we show in Fig. \ref{fig:hdb3} the comparison of gravitational
wave luminosities calculated with three different approaches.
F1 is calculated from $I_{ij}$ obtained by $x_i$, F2 is calculated
from $\dot{I}_{ij}$ obtained by $x_i$ and $\dot{x}_i$, and,
finally F3 is calculated from $\ddot{I}_{ij}$ obtained by
$x_i$, $\dot{x}_i$ and $\ddot{x}_i$. We can see that the differences
are smaller than factor 2 during almost the entire process.
Thus, we can conclud that the error concerning the numerical differentiation
is not a significant factor in our calculation. Hereafter, we adopt F3, which
is expected to cause the least error among the three methods.

\begin{figure}[t]
\includegraphics[width=77mm]{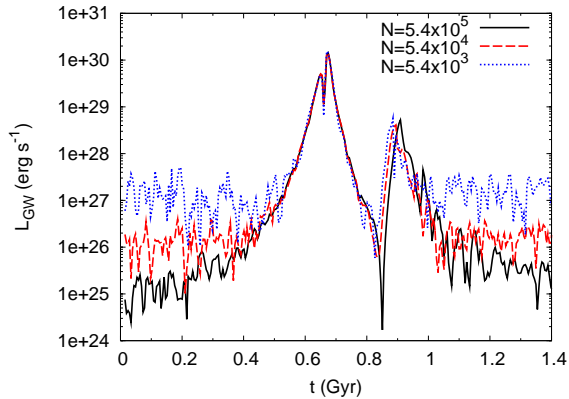}
\caption{Dependence on particle number ($5.4 \times 10^5$: solid,
$5.4 \times 10^4$: dashed, $5.4 \times 10^3$: dotted).
Other parameters are the same as the model-0.}
\label{fig:number}
\end{figure}

Next, we compare the luminosity evolution by varying the particle
number $N$ in Fig. \ref{fig:number}. As can be seen, however,
the two peaks are almost independent of the input particle number,
while the steady emission after and before collision shows
substantial increases for smaller $N$. Thus, we can expect
that the feature of the peaks and total emitted energy can
be studied reliably with particle numbers larger than $\sim 10^4$.
This particle number should be compared with that of
the simulations in the next subsection, $N = 2.4 \times 10^5$.
On the other hand, the steady emission is affected by
the resolution of simulation and would be highly suppressed
in real galaxy merger.

\begin{figure}[t]
\includegraphics[width=77mm]{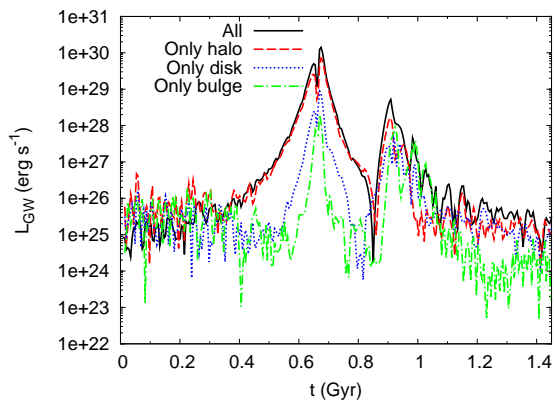}
\caption{Time evolution of luminosity of gravitational waves
for model-0
with contribution from halo (dashed), disk (dotted),
bulge (dash-dotted) only. Total luminosity is also shown (solid).}
\label{fig:hdb}
\end{figure}

Having established the above two elements of this model,
we next study the relative importance of galaxy components.
Although the mass of a galaxy is mostly due to the contribution 
from its halo, disk and bulge may also considerably 
contribute to gravitational wave emission
because they are more concentrated than the halo itself.
To account for this, we calculate the quadrupole $I_{ij}$, Eq. (\ref{Iij}),
for each component. However, the dynamics of particles in all
components was solved taking the gravitational potential of
all the particles into account.
Fig. \ref{fig:hdb} shows the contribution of each component
(halo, disk and bulge). 
We can see that, at the peaks, the luminosity is dominated by
the contribution from the halo and other contributions are
at most $12\%$. Remembering that the mass of the disk is
about one seventh of that of the halo itself, and that the luminosity is
apparently proportional to the squared mass. The contribution
of the disk ($\sim 10\%$) is larger than a naive expectation.
This would be due to the fact that the disk is more concentrated
than the halo, so that the disk emits gravitational waves
more effectively. We tested the effect of
concentration of the disk component by changing the mass profile
artificially. As a result, a disk with size $80\%$ and $120\%$ of
the fiducial size contributes $14\%$ and $9\%$ of the total
luminosity, respectively, while it is $12\%$ for the fiducial
mass profile. Thus, it is expected that the gravitational wave
luminosity is dominated by the halo component for a reasonable
range of the disk concentration.

Furthermore, we note that the ratio of the luminous mass to
the dark mass $M_{\rm dark}/M_{\rm lum}$ that we adopted is about 7,
which is smaller than that of typical galaxies ($\sim 10$).
Thus, our calculations are overestimating the disk and bulge
contributions. This fact justifies our later calculations
with galaxies obtained from a cosmological simulation which
have only halo components.

\begin{figure}[t]
\includegraphics[width=77mm]{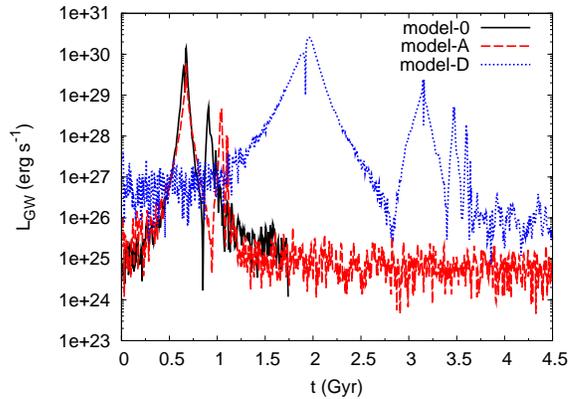}
\caption{
Time evolution of luminosity of gravitational waves
for model-0 (solid), model-A (dashed), and model-D (datted).}
\label{fig:GW-0AD}
\end{figure}

\begin{figure}[t]
\includegraphics[width=77mm]{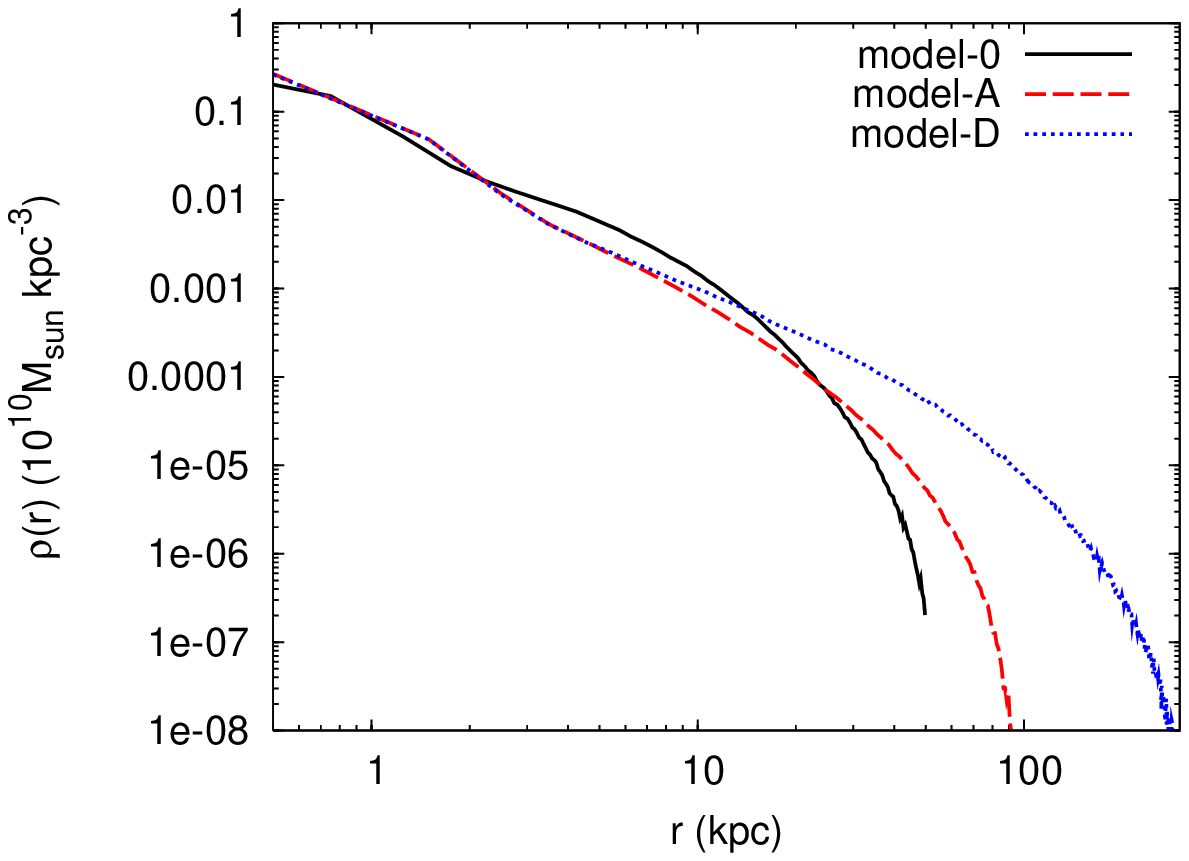}\\
\includegraphics[width=77mm]{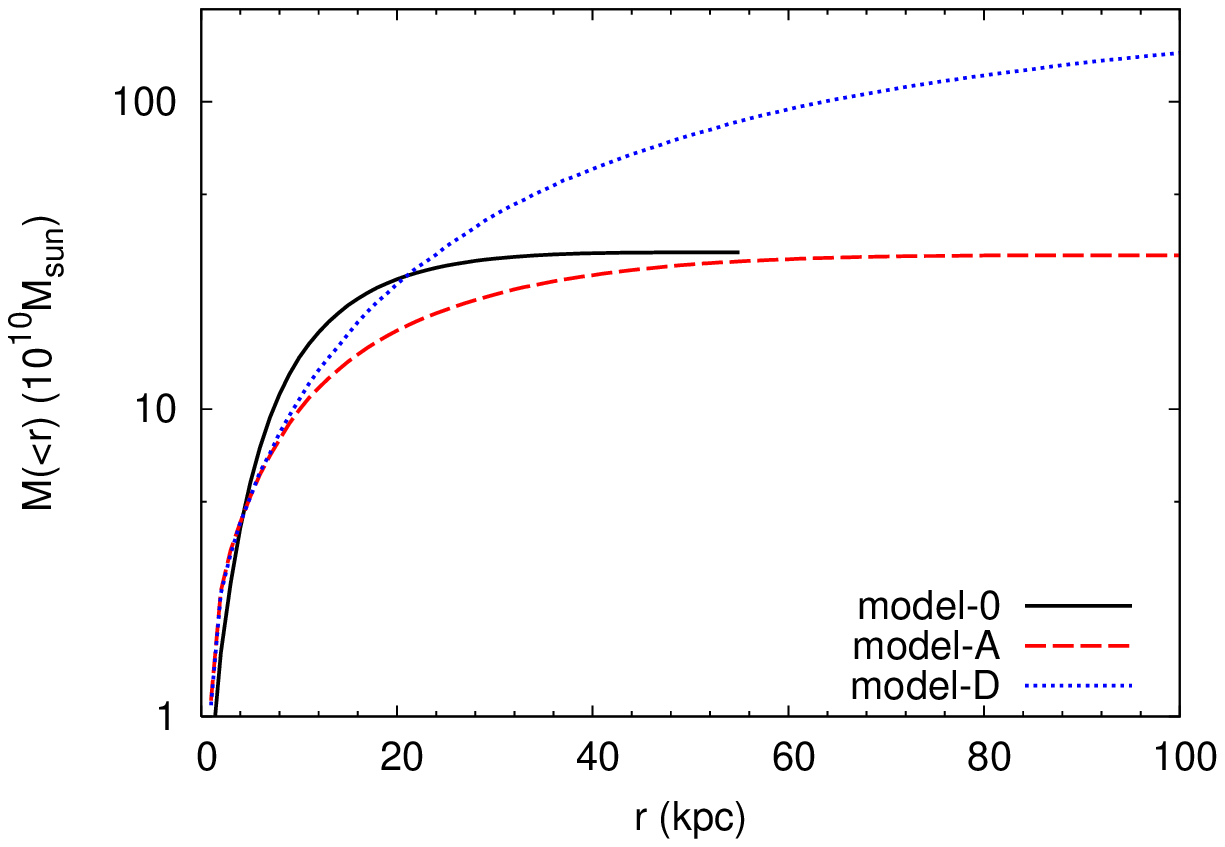}
\caption{
Top: the density profile of model-0 (solid), model-A (dashed), 
and model-D (dotted). 
Bottom: the mass of model-0 (solid), model-A (dashed), 
and model-D (dotted) within the radius.}
\label{fig:massprofile-0A}
\end{figure}

Finally, we compare the gravitational-wave luminosities
for mergers of model-0, model-A and model-D galaxies.
The peak luminosity for model-A is $4.7 \times 10^{29}~{\rm erg/sec}$,
which is about one third of that of model-0, even though both models
have almost the same mass. This is because model-0 has a more
concentrated halo than model-A as can be seen from Fig.
\ref{fig:massprofile-0A}. On the other hand, the peak luminosity
for model-D is $2.4 \times 10^{30}~{\rm erg/sec}$, 
which is approximately 5 times larger than that of model-A. 
Since the bulge and disk components of model-A and D are almost 
identical, the difference of peak luminosities comes from the halo mass 
components of these models. 
The halo mass distributions of the two models
are similar within 10 kpc, while model-D has more extended halo than 
model-A as is shown in the bottom panel of Fig. 10. 
Accordingly, the ratio of halo masses of model-D and A is about 7.6.
Thus, gravitational-wave luminosity is very sensitive to
both the mass and the density structure of halo, making quite clear
the importance of using a realistic model.
In the next subsection, we simulate mergers of galaxies
with a NFW-like density profile and study the mass dependence.

\subsection{Simulations with Type B Initial Conditions}
\label{NFW}

\begin{table}[t]
\caption{Models used in this subsection.
Column(1):model name.
Column(2):mass of one galaxy.
Column(3):initial relative velocity (relative velocity in
the $z$-direction is set to zero).
Column(4):initial separation between the galaxies
in the $x$-direction (initial separations
in the $y$- and $z$-directions are set to zero).}
\label{tbnfw}
\begin{tabular}{cccc}
\hline\hline
model & $M (h^{-1}M_{\odot}) $
& $(\Delta V_x, \Delta V_y)$ (${\rm km~s^{-1}}$)
& $\Delta x$ ($h^{-1}{\rm kpc}$) \\
\hline
A-1 & $3.8 \times 10^{12}$ & (0,0) & 1200 \\
A-2 & $3.8 \times 10^{12}$ & (220,0) & 1200 \\
A-3 & $3.8 \times 10^{12}$ & (380,0) & 1800 \\
B-1 & $3.8 \times 10^{12}$ & (0,0) & 850 \\
B-2 & $3.8 \times 10^{12}$ & (0,70) & 850 \\
B-3 & $3.8 \times 10^{12}$ & (0,140) & 850 \\
C-1 & $3.8 \times 10^{11}$ & (0,0) & 680 \\
C-2 & $3.8 \times 10^{10}$ & (0,0) & 680 \\
\hline
\end{tabular}
\end{table}

\begin{figure}[t]
\includegraphics[width=70mm]{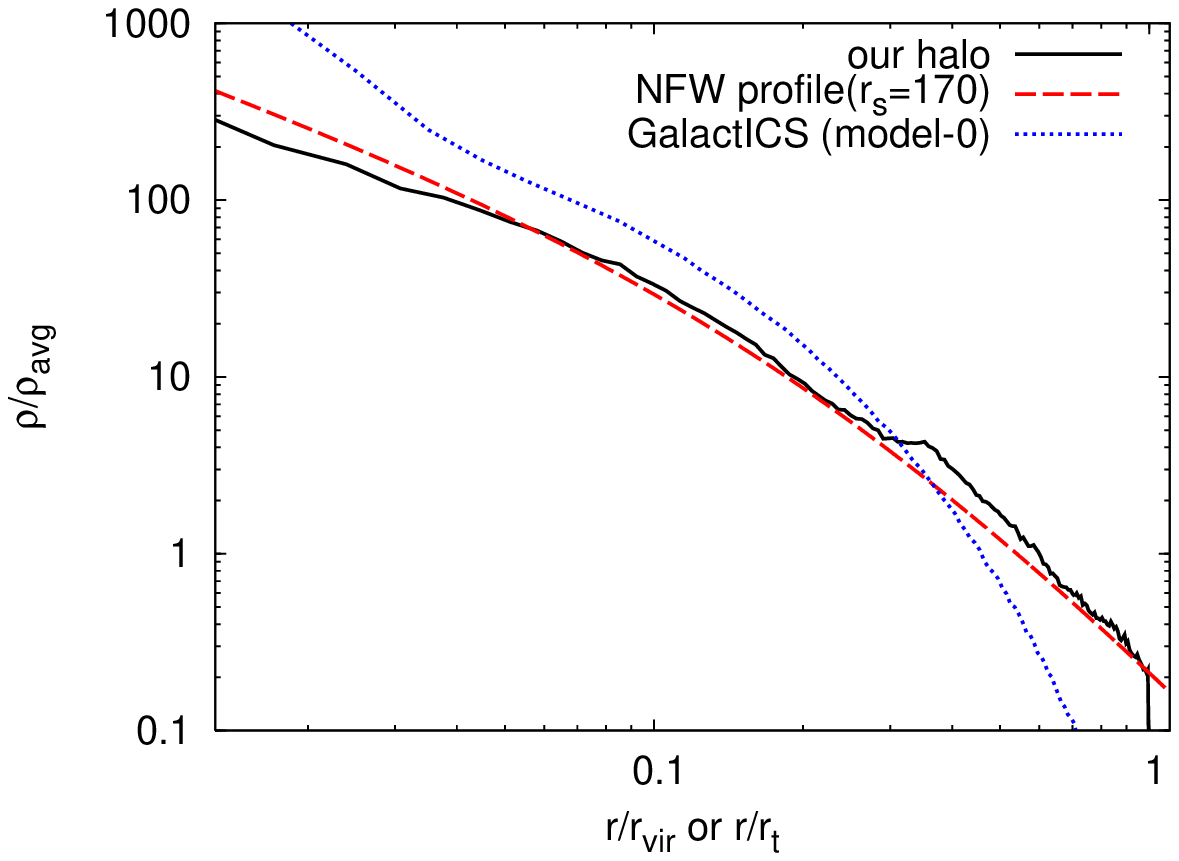}\\
\includegraphics[width=70mm]{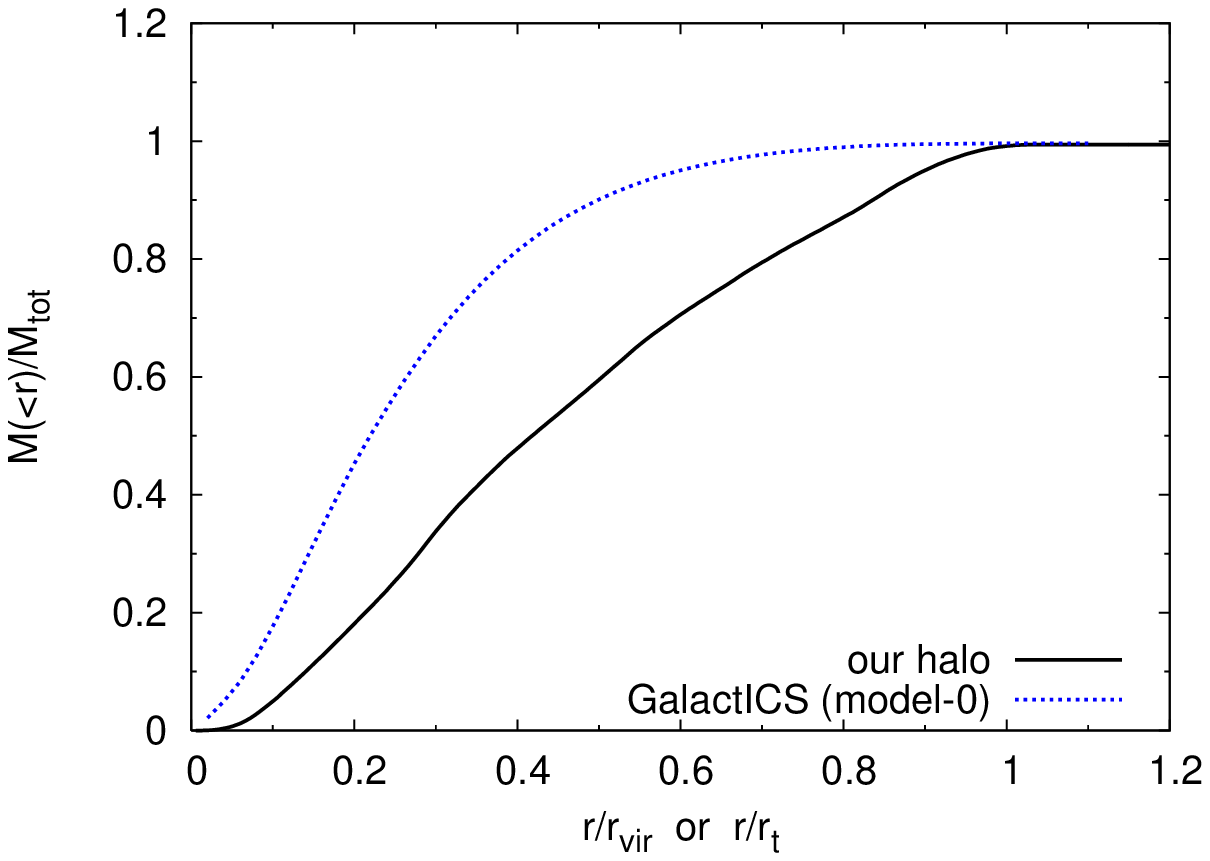}\\
\includegraphics[width=70mm]{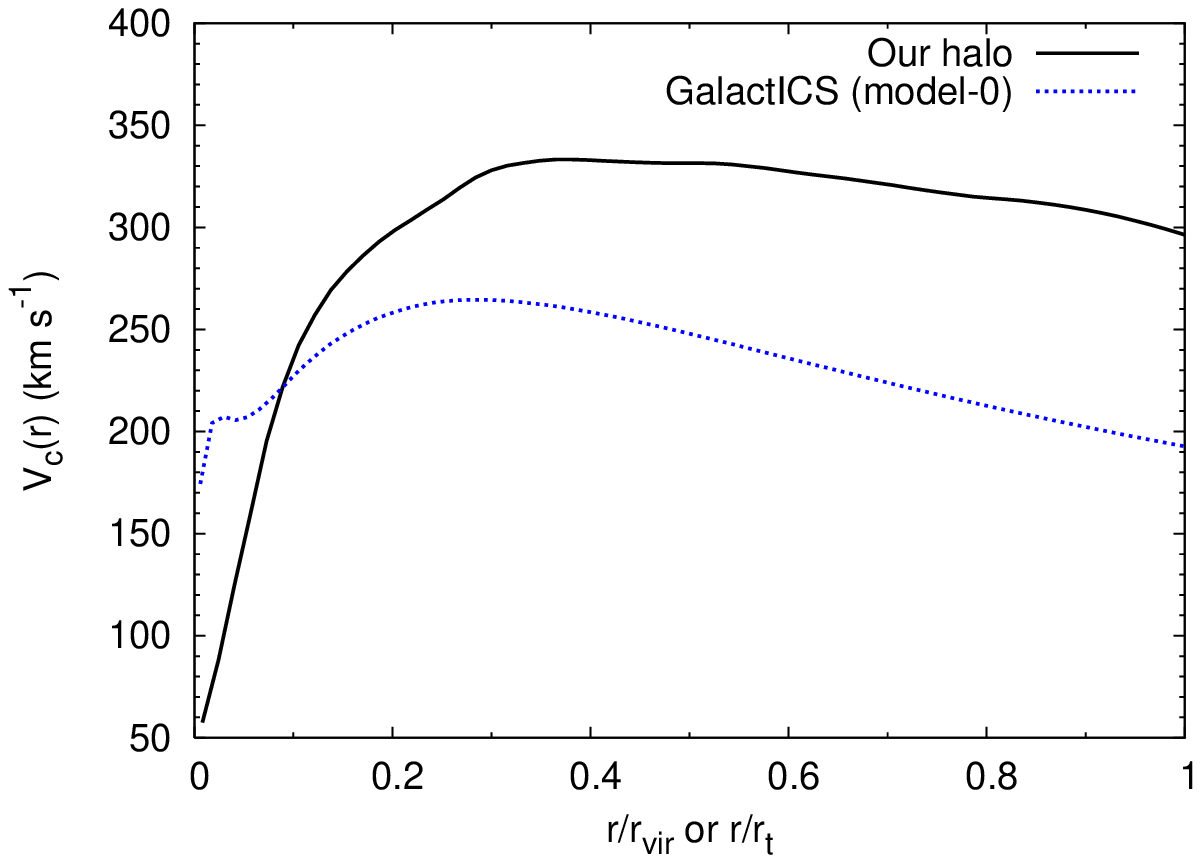}
\caption{
Top: Density profile of the fiducial halo with
$1.3 \times 10^{14} h^{-1}M_\odot$ (solid)
which is well fitted by the NFW profile with
$r_{\rm s} \sim 170~h^{-1}{\rm kpc}$ (dashed).
For reference, the density profile of the galaxy (model-0)
used in section \ref{basic} is also shown (dotted).
The density and the radius are normalized 
with the average density $\rho_{\rm avg}$, and
the virial radius $r_{\rm vir}$ of each galaxy.
However, the virial radius of the galaxy used in section \ref{basic} 
is replaced by the tidal radius.
Middle: the mass of the fiducial galaxy
(solid) and one used in section \ref{basic} (dotted)
within the radius. 
As in the top figure the radius is normalized, and 
the axis of ordinate is normalized by the total mass 
of each model.
Bottom: rotation curves of the fiducial galaxy
(solid) and one used in section \ref{basic} (dotted).}
\label{fig:fit}
\end{figure}

In this subsection, we adopt a dark halo from a cosmological
simulation (for details, see \cite{Masaki}) to prepare initial
conditions. 
The halo identification is done in a two-step
manner. First, we select candidate objects using the 
friends-of-friends (FOF) algorithm \cite{1985ApJ...292..371D}. 
We set the linking parameter $b=0.2$. 
Secondly, we apply the spherical over density algorithm to the 
located FOF groups. To each FOF group
we assign a mass such that the enclosed 
mass within the virial radius is $\Delta \times \rho_{\rm crit}(z)$, 
where $\rho_{\rm crit}(z)$ is the critical density 
for which the spatial geometry is flat. 
Based on the spherical collapse model \cite{1980lssu.book.....P}, 
$\Delta$ is 200. The adopted halo has mass
$1.3 \times 10^{14} h^{-1}M_\odot$, particle number
$\sim 2.4 \times 10^5$, and a density profile well approximated
by the NFW profile,
\begin{equation}
\rho_{(r)}
= \frac{\rho_{\rm s}}{(r/r_{\rm s}) (1+r/r_{\rm s})^2},
\end{equation}
where $r_{\rm s} \sim 170~h^{-1}{\rm kpc}$ (Fig. \ref{fig:fit}).
We placed two identical halos with various initial relative
velocities. Also we vary mass of one or both of two galaxies
to study the dependence of mass (ratio) 
by a scaling law based on
an empirical relation obtained in \cite{2001MNRAS.321..559B},
$M_{\rm vir} \propto V_{\rm max}^{\alpha}$, where $\alpha \sim 3.4$.
Here $L$, $V_{\rm max}$ and $M_{\rm vir}$ are luminosity,
maximum circular velocity and virial mass, respectively.
This relation is realistic taking into 
account galaxy rotation.
From this relation, the following relation is obtained
\cite{2005MNRAS.357..753G},
\begin{equation}
r' = \left( \frac{M'}{M_0} \right)^{1-2/\alpha} r_0,
\label{eq:TF}
\end{equation}
where $r_0$ and $M_0$ are the original radius and mass,
and $r'$ and $M'$ are the scaled radius and mass, respectively.

Table \ref{tbnfw} summarizes the models used for our calculations.
Models A-1, A-2 and A-3 are used to study the dependence of initial
relative velocity in the direction of initial separation
($x$-direction). Effects of initial relative velocity in
the $y$-direction, that is, the relative angular momentum,
are studied by models B-1, B-2 and B-3. Finally, models C-1
and C-2 have different mass ratio, $1/10$ and $1/100$, respectively.

\begin{figure}[t]
\includegraphics[width=77mm]{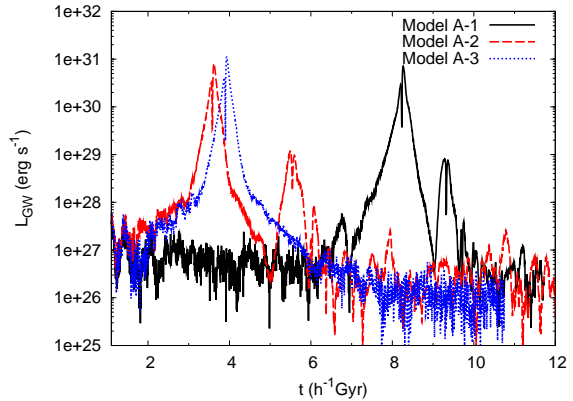}
\caption{Comparison of luminosity evolution varying the initial
relative velocity in the direction of initial separation
(models A-1, A-2 and A-3).}
\label{fig:nfw1}
\end{figure}

Fig. \ref{fig:nfw1} shows the comparison of luminosity
evolution varying the initial relative velocity in the
direction of initial separation (models A-1, A-2 and A-3).
Here, the maximum relative velocity with which
two galaxies are gravitationally bound is about
$220~{\rm km~sec^{-1}}$. Therefore, model A-2 represents a case
where two galaxies are marginally bound, while
model A-3 represents a case where two galaxies are not bound and pass through
each other after the first collision. This is why
there is only one peak in the luminosity of model A-3.
Other than this point, we cannot see any differences
between the three models in this figure that is,
the peak luminosity and width look almost independent of
$\Delta V_x$.

\begin{figure}[t]
\includegraphics[width=77mm]{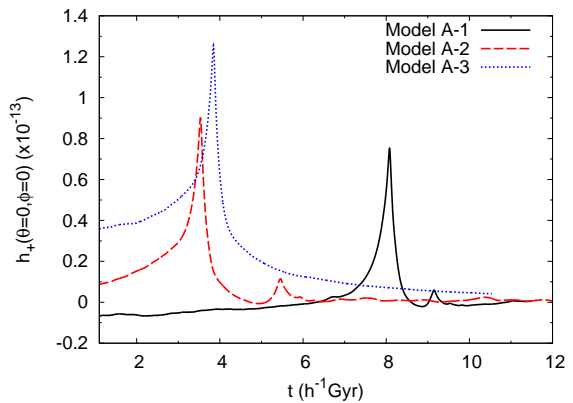}
\caption{Amplitude of the plus mode in the $z$-direction
at $10~h^{-1}{\rm Mpc}$ from the galaxies for models A-1, A-2
and A-3.}
\label{fig:h1plus}
\end{figure}

To see the difference between these three models more closely,
we plot the amplitude of the plus mode in the $z$-direction
in Fig. \ref{fig:h1plus}. The amplitude is calculated at
$10~h^{-1}{\rm Mpc}$ from the galaxies and the peak amplitude
is of order $10^{-13}$. The difference in the amplitudes is
about $20-50\%$ and increases as $\Delta V_x$ increases.

\begin{figure}[t]
\includegraphics[width=77mm]{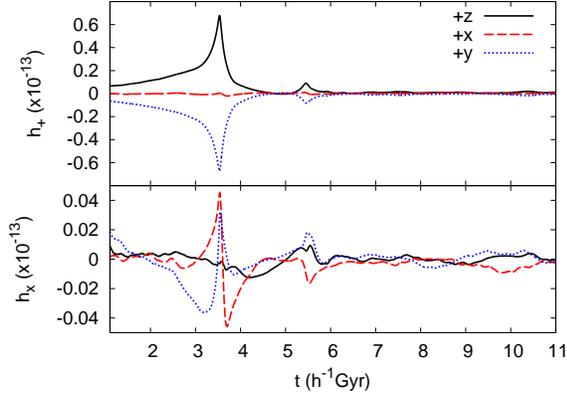}
\caption{Amplitude of the plus (top) and cross (bottom) modes 
in the $x$-, $y$- and $z$-directions 
at $10~h^{-1}{\rm Mpc}$ from the galaxies for model A-2.}
\label{fig:h12plus}
\end{figure}

Fig. \ref{fig:h12plus} 
shows the amplitude of the plus and cross modes in
the $x$-, $y$- and $z$-directions for model A-2.
We see that gravitational waves are mostly emitted as
the plus mode in $y$- and $z$-directions, and the cross mode
is smaller by one order. This comes from the fact that
the galaxy motion is almost axisymmetric with respect to
the $x$-axis. The small difference between radiation of
the $y$- and $z$-directions is due to the small
deviation of galaxy structure from the spherical symmetry.

\begin{figure}[t]
\includegraphics[width=77mm]{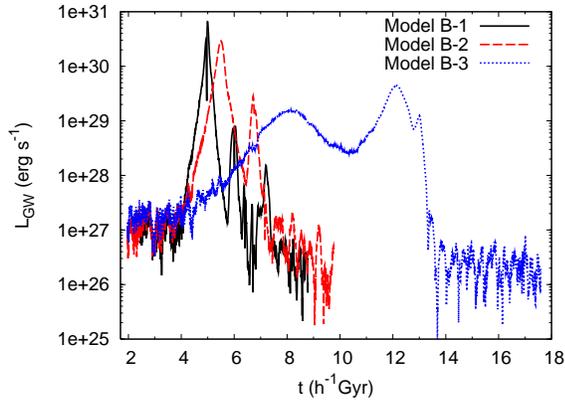}
\caption{Luminosity evolution varying the initial relative
velocity in the direction orthogonal to the direction of
the initial separation (models B-1, B-2 and B-3).}
\label{fig:elliptical}
\end{figure}

In Fig. \ref{fig:elliptical}, we show the comparison
of luminosity evolution varying the initial relative
velocity in the direction orthogonal to the direction of
the initial separation (models B-1, B-2 and B-3).
When the two galaxies have relative angular momentum,
it takes more time to collide and relax into a single
galaxy. Thus, the peak width broadens for large $\Delta V_y$
while the peak luminosity is suppressed.

\begin{figure}[t]
\includegraphics[width=77mm]{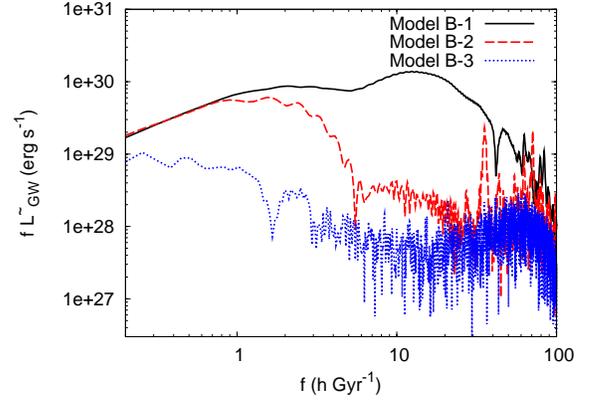}
\caption{Fourier spectrum of the luminosity for
Models B-1, B-2, B-3.}
\label{fig:FT}
\end{figure}

Fig. \ref{fig:FT} is the spectrum of the luminosity
for models B-1, B-2 and B-3. Here the spectrum is calculated
by the discrete Fourier transform,
\begin{equation}
\tilde{L}_{\rm GW}(f_n)
= \Delta t
  \sum_{k=0}^{N-1} L_{\rm GW}(t_k)
                   \exp\left(-\frac{2\pi ikn}{N}\right)
\end{equation}
where $n = 0, \cdots, N-1$ and $N$ is the number of time bins.
Here $\Delta t = 5/h~{\rm Myr}$ is the width of each time bin
and $t_k = k \Delta t$ is the time of the $k$-th bin.
Gravitational wave frequency is represented by
$f_n = n/(N \Delta t)$. We can see that increasing $\Delta V_y$
decreases the characteristic frequency which corresponds
to the peak width, as is expected from Fig. \ref{fig:elliptical}.

\begin{figure}[t]
\includegraphics[width=77mm]{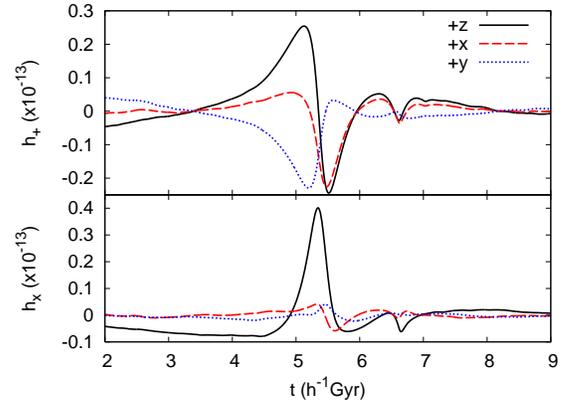}
\caption{Amplitude of the plus (top) and cross (bottom) modes 
in the $x$-, $y$- and $z$-directions 
at $10~h^{-1}{\rm Mpc}$ from the galaxies for model B-2.}
\label{fig:h12plus_B-2}
\end{figure}

Relative angular momentum also affects the direction dependence
of the emission. 
Fig. \ref{fig:h12plus_B-2} shows the amplitude of the plus and
cross modes in the $x$-, $y$- and $z$-directions for model B-2,
which correspond to Fig. \ref{fig:h12plus} for model A-2. 
We see that the plus and cross modes are emitted
with the same order of magnitude in the $z$-direction, while
there is no significant difference between model A-2 and B-2
in the $x$- and $y$-direction. This is because the galaxy motion
is no longer axisymmetric, due to the relative angular momentum.

\begin{figure}[t]
\includegraphics[width=77mm]{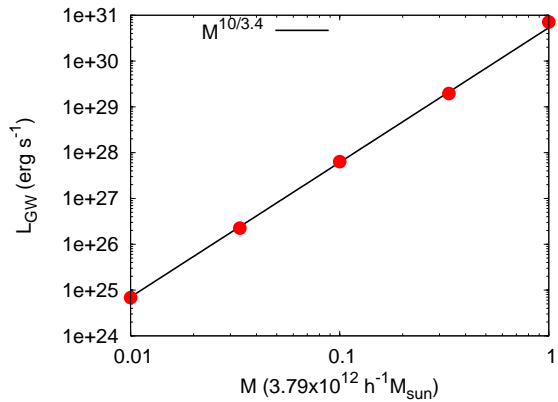}
\caption{Peak luminosity as a function of mass fixing
mass ratio to $1:1$.}
\label{fig:mass}
\end{figure}

Fig. \ref{fig:mass} shows the dependence of the peak luminosity
as a function of the galaxy mass $M$ fixing the mass ratio
to $1:1$. As can be seen, the luminosity is well fitted
by $M^{10/3.4}$. One may expect that the luminosity is
proportional to $M^2$ because the quadrupole is proportional
to $M$. However, this is not the case because the length and
time scales are also dependent on $M$. We can understand
this behavior as follows.

Defining the typical length and time scales as $L$ and $T$,
respectively, the quadrupole and luminosity are roughly given by,
\begin{equation}
I \sim M L^2, ~~~
\dddot{I} \sim \frac{M L^2}{T^3}, ~~~
L_{\rm GW} \sim \frac{M^2 L^4}{T^6}.
\end{equation}
On the other hand, assuming the gravitational potential energy
between the two galaxies is equal to the kinetic energy of
their relative motion at the collision, characteristic relative
velocity $V \sim L/T$ is evaluated as,
\begin{equation}
V^2 \sim 2 \frac{GM}{r}
    \sim \frac{M}{L}.
\end{equation}
Thus, we have,
\begin{equation}
T^2\sim\frac{L^3}{M},
\end{equation}
and then,
\begin{equation}
L_{\rm GW}
\sim \frac{M^2 L^4}{T^6}
\sim \frac{M^5}{L^5}.
\end{equation}
From Eq. (\ref{eq:TF}), we have $L \propto M^{1-2/\alpha}$ and
finally obtain,
\begin{equation}
L_{\rm GW} \propto M^{\frac{10}{\alpha}},
\end{equation}
with $\alpha \sim 3.4$.

\begin{figure}[t]
\includegraphics[width=77mm]{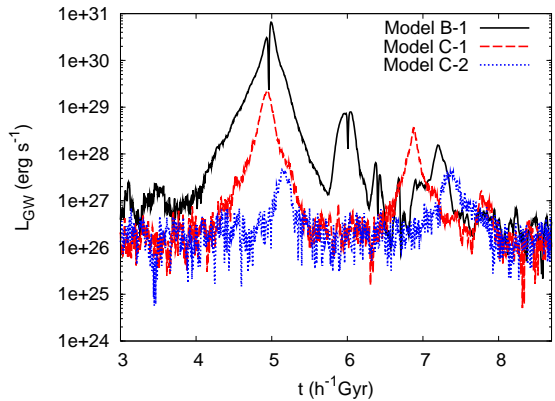}
\caption{Luminosity evolution varying the mass ratio
(models B-1, C-1 and C-2).}
\label{fig:ratio}
\end{figure}

Fig. \ref{fig:ratio} shows the comparison of luminosity
evolution varying the mass ratio. The peak luminosity
is roughly proportional to $(M_1/M_2)^{1.5}$ where
$M_1 < M_2$. This dependence may be understood by
replacing $M$ with $\sqrt{M_1 M_2}$ in the discussion
in the previous paragraph.

\section{Summary and Discussion}
\label{summary}

In this paper, we performed a systematic study of gravitational
waves from galaxy mergers through $N$-body simulations
with Gadget2. Two types of initial condition were adopted.
First we used galaxies generated by ``GalactICS'' which consist
of disk, bulge and halo. We showed that the features of
peaks such as the width and luminosity are reliably simulated
with particle numbers larger than $\sim 10^4$. Following this,
the relative importance of components was investigated
and it was found that the dominant contribution to
gravitational-wave luminosity comes from halo component, while 
contributions from other components were less than $10\%$.

Second, we used initial conditions with a dark halo adopted
from a cosmological simulation. This halo had a realistic
density structure and was more suitable for a precise estimation
of gravitational-wave luminosity. The peak luminosity amounted
to $10^{31}~{\rm erg~sec^{-1}}$ for the collision of two
halos with mass $3.8 \times 10^{12}~h^{-1}M_{\odot}$.
We showed that the initial relative velocity in the direction of the initial
separation does not significantly affect gravitational wave emission,
with a difference of about $20-50\%$ in amplitude.
On the other hand, we found that 
the initial relative angular momentum broadens
the peak width and suppresses the luminosity by one order.
Mass dependence of the peak luminosity was also investigated.
In contrast to the naive expectation, $L_{\rm GW} \propto M^2$,
we obtained $L_{\rm GW} \propto M^3$. We gave a simple
analytic interpretation of this behavior based on the scaling
relation of the mass and size of galaxies. Also this analysis
was shown to be applicable to the dependence on the mass ratio.

To argue the observability of gravitational waves from galaxy
merger through the B-mode polarization of CMB, it is necessary
to evaluate gravitational wave background, which is contributed to
by the entire history of galaxy merger. 
In future research we will present a detailed estimation
by combining the elementary process studied here and
the merger history based on hierarchical galaxy formation model
\cite{2004ApJ...615...19E}.
Here we give a very rough order-of-magnitude estimation
based on several simplifying assumptions. Let us assume
that all the current galaxies have mass of $10^{12} h^{-1} M_\odot$
and were created through successive mergers starting from
the progenitor galaxies with mass $10^{10} h^{-1}M_\odot$.
Further, we assume that all the mergers took place with mass ratio
1:1 and head-on orbit, and neglect the redshift of gravitational
waves. Counting the number of mergers in the current
horizon volume and multiplying the gravitational-wave energy
emitted at each merger taking the scaling law in Fig. \ref{fig:mass}
into account, we obtain the density parameter of gravitational wave
background as
$\Omega_{\rm GW} = \rho_{\rm GW}/\rho_{\rm c} \sim 10^{-21}$
where $\rho_{\rm c}$ is the critical density.
This corresponds to inflationary gravitational waves with
the tensor-to-scalar ratio $\sim 10^{-6}$. Although it would be
too small to be detected through B-mode polarization of CMB
by the existing instruments, delensing of CMB polarization
may allow us to probe such weak gravitational wave background
in the future \cite{2004PhRvD..69d3005S,2005PhRvL..95u1303S}.

\begin{acknowledgements}

TI appreciates Vicent Quilis, and
A. C\'esar Gonz\'alez-Garc\'{\i}a for the detailed information
on their simulation \cite{2007PhRvD..75j4008Q}.
This work is supported in part by JSPS
Grant-in-Aid for the Global COE programs, ``Quest for
Fundamental Principles in the Universe: from Particles to
the Solar System and the Cosmos'' at Nagoya University.
KT is supported by Grant-in-Aid for Scientific Research No.~21840028.
NS is supported by Grant-in-Aid for Scientific Research No.~22340056
and 18072004. 
This research has also been supported in part by World
Premier International Research Center Initiative, MEXT,
Japan.

\end{acknowledgements}

\bibliography{GW}

\end{document}